\documentclass[%
	reprint,
showpacs,
	amsmath,amssymb,
]{revtex4-1}
\usepackage{color}
\usepackage{graphicx}
\usepackage{subfigure}
\usepackage{dcolumn}
\usepackage{bm}
\graphicspath{{fig/}}

\begin{document}

\title{Extend the explanation of transformation optics in metamaterial-modified wireless power transfer system}

\author{Lin Zhu}
\affiliation{
Department of Physics and Astronomy, Key Laboratory of Artificial Structures and Quantum Control (Ministry of Education), Shanghai Jiao Tong University, Shanghai 200240, China
}

\author{Xudong Luo}%
\email{luoxd@sjtu.edu.cn}

\affiliation{
Department of Physics and Astronomy, Key Laboratory of Artificial Structures and Quantum Control (Ministry of Education), Shanghai Jiao Tong University, Shanghai 200240, China
}%


\author{Hongru Ma}
\affiliation{
School of Mechanical Engineering; The State Key Laboratory of Metal Matrix Composites; and Key Laboratory of Artificial Structures and Quantum Control (Ministry of Education), Shanghai Jiao Tong University, Shanghai 200240, China
}%
\date{\today}

\begin{abstract}
Based on rigorous scattering theory we establish a systematic methodology for research of metamaterial-modified current-carrying conductors, from which we mathematically demonstrate the explanation of transformation optics could be extended in metamaterial-modified wireless power transfer system, and based on that we could establish a equivalent model. More important, our demonstration reveals that the equivalent model will still be applicable even when TO could not give a direct explanation, as the requirements of complementary media is not satisfied. And numerical results from our methodology as well as COMSOL verified our findings. The demonstration is not under specific frequency, the conclusion could be extended to a broad range of wavelength, and expected to be applicable for active cloak etc.

\end{abstract}

\pacs{42.25.Fx, 41.20.Jb, 88.80.ht}
\maketitle

\section{Introduction}

Metamaterials \cite{sihvola2007metamaterials,shamonina2007metamaterials} are widely applied in many areas \cite{pendry2000perfectlens,metaWPT,antenna,experimentWPTmeta}; in particular, they have established the experimental foundation for transformation optics (TO) \cite{tranU,pendry2006controlling,chen2010transformation}. Moreover metamaterial-modified wireless power transfer (WPT) \cite{metaWPT,experimentWPTmeta,superlensWPT} are attractive for researchers, so as the development in active cloak devices \cite{PhysRevLett.100.063904active,ieee6330979active,PhysRevX.3.041011active}.

It is well known that TO works well for the manipulation of EM waves \cite{antenna,yang2008superscatterer,luo2009conceal,metaharvesting,compactShort}. Most of the time TO will give a direct geometric explanation, which may greatly simplify the explanation for metamaterial's modification. While in metamaterial modified WPT, the TO is not well applied, as WPT system are based on current-carrying conducting wire and power transfer, which means those results based on perfect electric conductor (PEC) could not be applied directly in WPT systems. Rather than numerical analysis from finite element method (FEM), to establish rigorous TO explanation in such system, an analytical analysis is essential. What's more, in WPT or active cloak researches, the objects (e.g. receiver) might destroy the condition the TO explanation will rely on. As a result it is hard to extend TO explanation in such situation.

As a matter of fact, we develop a general methodology based on rigorous scattering theory to address metamaterial-modified current-carrying conductors, from which we are able to discuss the TO explanation mathematically. We choose super-scatterer \cite{yang2008superscatterer} as the metamedia to modify the WPT. To clarify our conclusions we discuss a well simplified WPT system, with only two parallel conducting wires acting as an receiver and emitter, we call it PCW in this article. Then we cover the emitter wire with super scatterer shell (ss-shell), we name it ss-PCW. With introduced applied voltage, we are able to discuss its power transfer properties. Based on this heuristic model and our methodology, we could establish an analytical equivalent model, which is extended from TO. It will greatly simplify the analysis of the ss-PCW, as it could avoid solving for detail EM fields in metamaterial, when we calculate the transfer power and efficiency of this system. Moreover, from our demonstration we find the equivalent model could still be applicable even when the requirements of complementary media\cite{pendry2003complementary} is not satisfied. Which greatly expand the usage scope of the equivalent model.

We started from discussing metamaterial modified current-carrying conductor because it is a heuristic model for WPT and active cloak. Although the results are obtained from the two-dimensional(2D) circular cylindrical case, it is expected to be applicable for a more general cases, even three-dimensional (3D) models. And our analytical demonstration is not under any certain frequency, the conclusion could be applied to a broad range of frequencies. Thus the deduction from our methodology could be easily extended to many other research fields.

This article is organized as follows: First, we consider the EM radiation and scattering problems for a super-scatterer shell (ss-shell)-covered conducting wire, in which current is introduced by impedance boundary conditions, where TO could be well extended. Second, we study a more complicated model, the ss-PCW. From which we give analytical analysis to discuss the limitation and extension of TO, and further analysis reveal that the equivalent model is still applicable even when TO could not be applied directly. Finally, we verify our demonstration through numerical results obtained from FEM and our series expansion solution.

\section{TO explanation for ss-shell covered conducting wire}

\begin{figure}
	\includegraphics[width=0.5\textwidth]{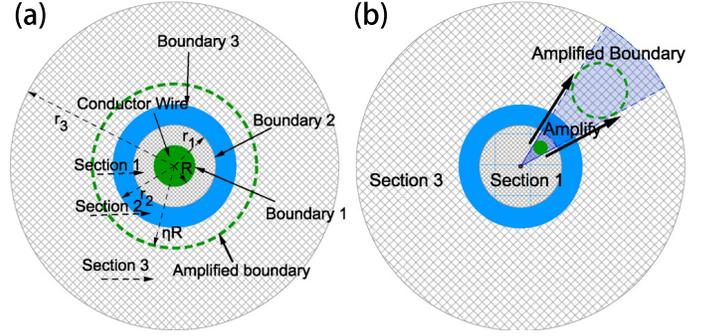}
	\caption{Simple schematic of a metamaterial-modified current-carrying conducting wire: (a),ss-shell covered conducting wire. (b),The conducting wire and ss-shell are not concentric.}
	\label{fig:simpleModel}
\end{figure}

To present TO explanation on an metamaterial modified wireless power transfer system, we begin with a simple but heuristic model of an ss-shell-covered conductor, as shown in Figure.\ref{fig:simpleModel}(a). It is a 2D model, and we divide it into three sections: section 1 ($0<r<r_1$), section 2 ($r_1<r<r_2$) and section 3 ($r_2<r<r_3$).In cylindrical coordinates, as shown in Refs.\cite{pendry2006controlling,yang2008superscatterer}, each section has different permittivity and permeability tensors,
\begin{equation}
	\left\{ {\begin{array}{*{20}{c}}
		{{\varepsilon _r} = {\mu _r} = \frac{{f(r)}}{r}\frac{1}{{f'(r)}}},\\
		{{\varepsilon _\theta } = {\mu _\theta } = \frac{r}{{f(r)}}f'(r)},\\
		{{\varepsilon _z} = {\mu _z} = \frac{{f(r)}}{r}f'(r)},
	\end{array}} \right.
	\label{eq:parametersFr}
\end{equation}
which can be derived from the following coordinate transformations:
\begin{equation}
	f(r) = \left\{ {\begin{array}{*{20}{l}}
		f_1(r)={\eta r }, \quad\quad\quad\quad {0 < r < {r_1},}\\
		f_2(r)={\eta {r_1} + \frac{({r^m} - {r_1^m})( \eta {r_1} - {r_2})}{{r_1^m} - {r_2^m}}},
		\quad {r_1} < r < {r_2}, \\
		f_3(r)=r, \quad\quad\quad\quad {r_2} < r < {r_3}.
	\end{array}} \right.
	\label{eq:frValue}
\end{equation}
where $m\ne 0 $.

From the viewpoint of TO (complementary media), section 3 (vacuum) will be complementary to section 2, which is an ss-shell filled with negative index material (NIM), and we will ultimately detect an amplified image originating from section 1 in the domain $0<r<r_3$.

This result can also be proven using the scattering theory of EM waves. As demonstrated in Ref.\cite{yang2008superscatterer}, if we consider only a transverse electric (TE)-polarized EM field with a harmonic time dependence of $e^{-i{\omega}t}$in our 2D model, the general series solution for the electric field (and the magnetic field) can be written as
\begin{equation}
	{E_z}(r,\theta ) = \sum\limits_{n =  - \infty }^\infty  {[{a_n}{J_n}({k_0}f(r)) + {b_n}H_n^{(2)}({k_0}f(r))]{e^{in\theta }}},
	\label{eq:2dEM}
\end{equation}
where $k_0$ is the wave vector in vacuum and $J_n$ and $H_n^{(2)}$ are the nth-order Bessel function and the nth-order Hankel function of the second kind, respectively.

As mentioned above, we can consider that any EM boundary inside the ss-shell is amplified from the domain $0 < r < {r_1}$ to $0 < r < {r_3}$, with an amplification factor of $\eta=\frac{r_3}{r_1}$. If we place a conducting wire of radius R in section 1, it will form an enlarged image with an amplified boundary in the domain $0 < r < {r_3}$, as shown in Figure.\ref{fig:simpleModel}.We typically consider a PEC surface when designing illusionary optical devices; here, however, we consider the general case of a current-carrying conducting wire (to which a time-varying voltage is applied). Before we solve for the parameters of the series expansion solution, we should apply impedance boundaries to this model. As discussed in Ref.\cite{yuferev1999selection}, the conductor we consider here has high conductivity, and the wavelength is also sufficiently large to satisfy the impedance boundary approximation. In fact, it is also not essential to solve the Maxwell equations inside the conductor to obtain the exact solution for the EM fields in that region; as a result, applying impedance boundary conditions could simplify our calculations.

Using the Maxwell equations and impedance boundary conditions, we can write the boundary condition for boundary 1 (the interface between the conducting wire and section 1 in Figure.\ref{fig:simpleModel}(a) ) as follows,
\begin{subequations}
	\begin{equation}
		{\frac{{{H_{\theta n}^c}(r)}}{{{E_{zn}^c}(r) - {E_{szn}^c}}}\left| {_{r = R}} \right. =  - \sqrt {\frac{{{\varepsilon _c} - j\frac{{{\sigma _c}}}{\omega }}}{{{\mu _c}}}} },
		\label{eq:bd1-1}
	\end{equation}
	\begin{equation}
		{\frac{{\partial_r {E_{zn}^c}(r)}}{{{H_{\theta n}^c }(r)}}\left| {_{r = R}} \right. =  - i\omega {\mu _{\theta 1} }},
		\label{eq:bd1-2}
	\end{equation}
	\label{eq:bd1}
\end{subequations}
where $n=0,\pm 1,\pm 2,\dots$, and the parameters ${\varepsilon _c},{\mu_c},{\sigma_c}$, and ${E_{szn}^c}$ denote the material permittivity, permeability, and conductivity and the induced electric field (induced by the applied AC power source), respectively. Moreover, $\mu_{\theta 1}$ denotes the $\theta$ components of the permeability tensors of the materials in section 1, which can be evaluated from equations (\ref{eq:parametersFr}).

In each section, the series expansion solution for the EM fields can be expressed as
\begin{footnotesize}
	\begin{subequations}
		\label{eq:emSection}
		\begin{eqnarray}
			{E_z^\alpha(r,\phi )} &&= \sum\limits_{n =  - \infty }^\infty  {E_{zn}^\alpha(r)} {e^{in\phi }}\nonumber\\ &&= \sum\limits_{n =  - \infty }^\infty  {[a_n^\alpha J_n({k_0}{f_\alpha}(r)) + b_n^\alpha{H_n^{(2)}}({k_0}{f_\alpha}(r))]{e^{in\phi }}},
			\label{eq:esection}
		\end{eqnarray}
		\begin{eqnarray}
			{H_\theta ^\alpha(r,\phi )} &&= \sum\limits_{n =  - \infty }^\infty  {H_{\theta n}^\alpha(r)} {e^{i n\phi }}, \nonumber\\ {\hbox{with}\;\;}{H_{\theta n}^\alpha(r)} &&={\frac{{{f'_\alpha}(r)[a_n^\alpha{J'_n}({k_0}{f_\alpha}(r))
+ b_n^\alpha {{H_n^{(2)}}'}({k_0}{f_\alpha}(r))]}}{{ - i\omega {\mu _{\theta \alpha}}}}},
			\label{eq:msection}
		\end{eqnarray}
	\end{subequations}
\end{footnotesize}
where the upper or lower index $\alpha$ is equal to $1,2$ or $3$ and denotes the corresponding $f_{\alpha}(r)$, $\mu_{\theta \alpha}$, $a_n^\alpha,b_n^\alpha$, $E_z^\alpha$ and $H_\theta^\alpha$ associated with section $1$, $2$ or $3$, respectively. In this manner, we can obtain the following three boundary equations for boundary 1 to boundary 3:
\begin{widetext}
	\begin{subequations}
		\begin{equation}
			\left\{ {\begin{array}{*{20}{c}}
				{[a_n^1{J_n}({k_0}{f_1}(r)) + b_n^1{H_n^{(2)}}({k_0}{f_1}(r))]\left| {_{r = R}} \right. = {E_{zn}^c}(R)},\\
				{\frac{{{f'_1}(r)[a_n^1{J'_n}({k_0}{f_1}(r)) + b_n^1{{H_n^{(2)}}'}({k_0}{f_1}(r))]}}{{ - i\omega {\mu _{\theta 1}}}}\left| {_{r = R}} \right. =  - ({E_{zn}^c}(R) - {E_{szn}^c})\sqrt {\frac{{{\varepsilon _c} - j\frac{{{\sigma _c}}}{\omega }}}{{{\mu _c}}}} },
			\end{array}} \right.
			\label{eq:simpleModel-EMboundary1}
		\end{equation}

		\begin{equation}
			\left\{ {\begin{array}{*{20}{c}}
				{[a_n^1{J_n}({k_0}{f_1}(r)) + b_n^1{H_n^{(2)}}({k_0}{f_1}(r))]\left| {_{r = {r_1}}} \right. = [a_n^2{J_n}({k_0}{f_2}(r)) + b_n^2{H_n^{(2)}}({k_0}{f_2}(r))]\left| {_{r = {r_1}}} \right.},\\
				{\frac{{{f'_1}(r)[a_n^1{J'_n}({k_0}{f_1}(r)) + b_n^1{{H_n}^{(2)}}' ({k_0}{f_1}(r))]}}{{ - i\omega {\mu _{\theta 1}}}}\left| {_{r = {r_1}}} \right. = \frac{{{f'_2}(r)[a_n^2{J'_n}({k_0}{f_2}(r))
+ b_n^2{{H_n}^{(2)}}' ({k_0}{f_2}(r))]}}{{ - i\omega {\mu _{\theta 2}}}}\left| {_{r = {r_1}}}, \right.}
			\end{array}} \right.
			\label{eq:simpleModel-EMboundary2}
		\end{equation}

		\begin{equation}
			\left\{ {\begin{array}{*{20}{c}}
				{[a_n^2{J_n}({k_0}{f_2}(r)) + b_n^2{H_n^{(2)}}({k_0}{f_2}(r))]\left| {_{r = {r_2}} = [a_n^3{J_n}({k_0}{f_3}(r)) + b_n^3{H_n^{(2)}}({k_0}{f_3}(r))]\left| {_{r = {r_2}}} \right.} \right.},\\
				{\frac{{{f'_2}(r)[a_n^2{J'_n}({k_0}{f_2}(r)) + b_n^2{{H_n}^{(2)}}' ({k_0}{f_2}(r))]}}{{ - i\omega {\mu _{\theta 2}}}}\left| {_{r = {r_2}} = \frac{{{f'_3}(r)[a_n^3{J'_n}({k_0}{f_3}(r)) + b_n^3{{H_n}^{(2)}}' ({k_0}{f_3}(r))]}}{{ - i\omega {\mu _{\theta 3}}}}\left| {_{r = {r_2}}}, \right.} \right.}
			\end{array}} \right.
			\label{eq:simpleModel-EMboundary3}
		\end{equation}
	\end{subequations}
\end{widetext}
where $n=0,\pm 1,\pm 2,\dots$. In theory, we can obtain all of the EM fields by solving the above equations obtained from the boundary conditions.

It is meaningful to study the EM fields for $r=\eta R$, considering that this position is simply the amplified boundary of the conducting wire from TO theory, as shown in Figure.\ref{fig:simpleModel}(a). Here, we deduce from equation (\ref{eq:frValue}) that $f_{1}(R)=f_3(\eta{R})=\eta{R}$, $f_{1}(r_1)=f_{2}(r_1)=r_3$ and $f_{2}(r_2)=f_{3}(r_2)=r_2$, all of which are easy to demonstrate. As a result, we have equations $\frac{{f_\alpha}'(r)}{\mu_{\theta{\alpha}}}=\frac{{f_\alpha}(r)}{r}$, and aforementioned equations  (\ref{eq:simpleModel-EMboundary1}), (\ref{eq:simpleModel-EMboundary2}) and (\ref{eq:simpleModel-EMboundary3}) above can be simplified as follows:
\begin{equation}
	\left\{ {\begin{array}{*{20}{c}}
		{a_n^1 = a_n^2 = a_n^3},\\
		{b_n^1 = b_n^2 = b_n^3},
	\end{array}} \right.
	\label{eq:simpleModel-all}
\end{equation}
where $n=0,\pm 1,\pm 2,\dots$.

Moreover, by comparing equation (\ref{eq:simpleModel-EMboundary1}) with equations (\ref{eq:emSection}), we ultimately obtain
\begin{subequations}
	\label{eq:simpleModel-amplifiedBoundary}
	\begin{eqnarray}
		E_{zn}^3(\eta R) &&= [a_n^3{J_n}({k_0}{f_3}(r)) + b_n^3{H_n^{(2)}}({k_0}{f_3}(r))]\left| {_{r = \eta R}} \right. \nonumber\\&&= E_{zn}^c(R),
		\label{eq:sM-ae}
	\end{eqnarray}
	\begin{eqnarray}
		H_{\theta n}^3(\eta R) &&= \frac{{{f'_3}(r)[a_n^3{J'_n}({k_0}{f_3}(r)) + b_n^3{{H_n}^{(2)'}}({k_0}{f_3}(r))]}}{{ - i\omega {\mu _{\theta 3}}}}\left| {_{r = \eta R}} \right. \nonumber\\&& = \frac{{ - (E_{zn}^c(R) - E_{szn}^c)\sqrt {\frac{{{\varepsilon _c} - j\frac{{{\sigma _c}}}{\omega }}}{{{\mu _c}}}} }}{\eta },
		\label{eq:sM-am}
	\end{eqnarray}
\end{subequations}
where $n=0,\pm 1,\pm 2,\dots$. These equations imply that there is an impedance boundary at $r=\eta R$,
\begin{equation}
	\frac{{H_{\theta n}^3(r)}}{{(E_{zn}^3(r) - E_{szn}^c)}}\left| {_{r = \eta R}} \right. = \frac{{ - \sqrt {\frac{{{\varepsilon _c} - j\frac{{{\sigma _c}}}{\omega }}}{{{\mu _c}}}} }}{\eta }.
	\label{eq:simpleModel-ampBoundary}
\end{equation}
Indeed, if we take
\begin{equation}
	\sigma ' = \eta^{-2}\sigma + i \omega {\varepsilon _c} (\eta^{-2}-1),
	\label{eq:amp-conductivity}
\end{equation}
equation (\ref{eq:simpleModel-ampBoundary}) will have the same form as equation (\ref{eq:bd1-1}). In other words, we can obtain an enlarged current-carrying conducting wire, as indicated by TO theory.

The equivalence is discussed in detail for the following two cases. If there are observers at $r>\eta R$, they will detect the same EM fields from the following two emitters:
\begin{enumerate}
\item[ a)] A conducting wire with radius $R$, conductivity $\sigma$ and applied voltage $E_{szn}^c$ that is covered by an ss-shell as shown in Figure.\ref{fig:simpleModel}(a).
\item[ b)] A conducting wire with radius $\eta R$, the conductivity $\sigma^\prime$ defined in equation (\ref{eq:amp-conductivity}) and applied voltage $E_{szn}^c$.
\end{enumerate}
Based on this equivalence, we refer to case b as the materialized model of case a later in this article.

All of the deductions above can also be used to explain the model depicted in Figure.\ref{fig:simpleModel}(b). However, it should be remarked that the TO method may be not applicable under certain circumstances, as we will discuss later in this article.

\section{Limitation and extension of TO in ss-PCW}
\subsection{EM field in ss-PCW}
\begin{figure}
	\includegraphics[width=0.5\textwidth]{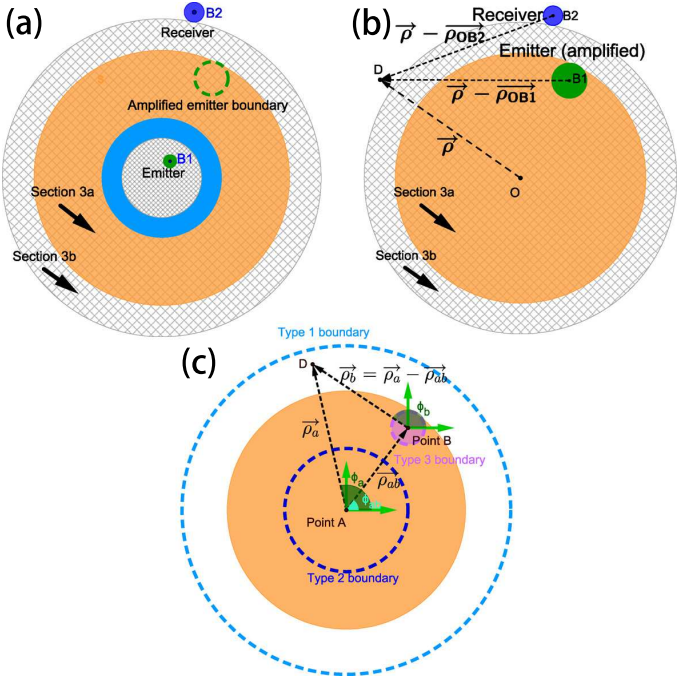}
	\caption{Schematic illustrations for ss-PCW:(a), a simplified WPT system with its emitter covered by an ss-shell, receiver could be located in any positions out of section 2 and 1. (b),materialized model for ss-PCW. (c),translation of wave functions in different coordinates. }
	\label{fig:sspcwschematic}
\end{figure}

Now that we have introduced TO explanation in the ss-shell-covered conducting wire model from rigorous scattering theory, we will discuss the ss-PCW, schematically illustrated in Figure.\ref{fig:sspcwschematic}(a).

The model can also be represented as a 3D model. An ss-shell-covered emitter wire forms part of a circuit (so are the receiver wire) with an time-varying applied voltage, which is equivalent to introduce a time-varying electromotive force by applying an external EM filed to the emitter circuit \cite{kurs2007science}. When electric energy is transferred from the emitter to the receiver, it can be regarded as a radiation and scattering problem of EM waves. For convenience in calculation, all conducting wires are circular cylinders; thus, that our previously applied methodology can also be used here as a suitable tool for calculating the transfer power and efficiency.

In the calculation, different receiver positions require different series expansion solutions. Unlike Figure.\ref{fig:simpleModel}, we divide section 3 into two parts: section 3a and section 3b. Both of them are in vacuum; section 3a is the domain with its radius smaller than the minimal radius which could cover the amplified emitter boundary. We will discuss the difference for receiver wire inside section 3a or outside of section 3a later.

Let us place the emitter and receiver wires in sections 1 and 3, respectively. As in equation (\ref{eq:2dEM}) above, we can write the EM fields in sections 1 to 3 as follows:
\begin{footnotesize}
	\begin{subequations}
		\label{eq:wpt-ss-shell}
		\begin{eqnarray}
			E_z^{(1)}({\rho},{\phi}) =&&
			\sum\limits_{n =  - \infty }^\infty [ {a_n^1{J_n}({k_0}f_{1}(|\overrightarrow {{\rho}}|)){e^{in{\phi}}}}   \\&&
				+ {b_n^1{H_n^{(2)}}({k_0}f_{1}(|\overrightarrow {{\rho}}-\overrightarrow {{\rho_{OB1}}}  |)){e^{in{\phi_{B1}}}}}]\nonumber
		\label{eq:wpt-ss-shell-1}
		\end{eqnarray}
		\begin{eqnarray}
			{E_z^{(2)}}(\rho ,\phi ) = && \sum\limits_{n =  - \infty }^\infty  [{d_n^2{J_n}({k_0}f_{2}(|\overrightarrow {{\rho}} |))}\\&&
				+  {g_n^2{H_n^{(2)}}({k_0}f_{2}(|\overrightarrow {{\rho}} |))]{e^{in{\phi}}}} ,\nonumber
			\label{eq:wpt-ss-shell-2}
		\end{eqnarray}
		\begin{eqnarray}
			E_z^{(3)}({\rho},{\phi}) =&& \sum\limits_{n =  - \infty }^\infty [ {a_n^3{H_n^{(2)}}({k_0}f_{3}(|\overrightarrow {{\rho}} |)){e^{in{\phi}}}}  \\&&
				+ {b_n^3{H_n^{(2)}}({k_0}f_{3}(|\overrightarrow {{\rho}}-\overrightarrow {{\rho_{OB2}}} |)){e^{in{\phi_{B2}}}}}], \nonumber
			\label{eq:wpt-ss-shell-3}
		\end{eqnarray}
	\end{subequations}
\end{footnotesize}
Here, we present only the expressions for the electric field; the corresponding $H_\theta$ components can be derived using the equation ${H_\theta } =  - \frac{1}{{i\omega \mu }}\frac{{\partial {E_z}}}{{\partial r}}$.

It is important to emphasize that the electric field (in sections 1 and 3) consists of three parts, each of which has a series expansion solution in different cylindrical coordinates. As shown in figures \ref{fig:sspcwschematic}(a) and \ref{fig:sspcwschematic}(b), point O is the center of the ss-shell, whereas in sections 1 and 3, $B\alpha$ is the center points of the emitter or receiver wires (where $\alpha=1,2$). In section 1 or 3, the origin points O, and $B\alpha$ offer two different cylindrical coordinate systems.

Following Weng Cho Chew' s book \cite{chew1995waves}, we present the translation of the wave functions corresponding to origin points A and B as follows (shown in Figure.\ref{fig:sspcwschematic}(c)):
\begin{footnotesize}
	\begin{eqnarray}
		&&H_m^{(2)}({k_0}|\overrightarrow {{\rho _b}} |){e^{im{\phi _b}}} = H_m^{(2)}({k_0}|\overrightarrow {{\rho _a}}  - \overrightarrow {{\rho _{ab}}} |){e^{im{\phi _b}}}
		\nonumber\\&&=
		\begin{cases}
			\sum\limits_{n =  - \infty }^\infty  H_{n - m}^{(2)}({k_0}\overrightarrow {{\rho _{ab}}} ){J_n}({k_0}\overrightarrow {{\rho _a}} ){e^{in{\phi _a} - i(n - m){\phi _{ab}}}}, &({\rho _a} < {\rho _{ab}}) 
			\\
			\sum\limits_{n =  - \infty }^\infty  {J_{n - m}}({k_0}\overrightarrow {{\rho _{ab}}} )H_n^{(2)}({k_0}\overrightarrow {{\rho _a}} ){e^{in{\phi _a} - i(n - m){\phi _{ab}}}}, &({\rho _a} > {\rho _{ab}} )
		\end{cases}
		\label{eq:waveFunctionTransform-All}
	\end{eqnarray}
\end{footnotesize}
where $m=0,\pm 1,\pm 2, \dots$. These equations describe the translation from origin point B to origin point A, and similar equations can be used for translation between any two of the three points, and similar for $J_{n}$. To solve for the expansion parameters of these series solutions, we need to apply the EM-field boundary conditions at each boundary.

Based on the translation equations given above, as previously discussed, we can write the boundary conditions as (with $n, m=0,\pm 1,\pm 2, \dots$)
\begin{widetext}
	\begin{subequations}

		\begin{equation}
			\small{
				\left\{ {\begin{array}{*{20}{c}}
					{{J_n}({k_0}\eta {R_1})\sum\limits_{m =  - \infty }^\infty  {[a_m^1{J_{m - n}}({k_0}\eta {r_{01}}){e^{ - i(n - m){\phi _1}}}]}  + b_n^1{H_n^{(2)}}({k_0}\eta {R_1}) = E_{zn}^{(1)}({R_1})},\\
					{{J'_n}({k_0}\eta {R_1})\sum\limits_{m =  - \infty }^\infty  {[a_m^1{J_{m - n}}({k_0}\eta {r_{01}}){e^{ - i(n - m){\phi _1}}}]}  + b_n^1{H_n^{(2)}}'({k_0}\eta {R_1}) = \frac{{i\omega }}{\eta }\sqrt {\frac{{{\varepsilon _1} - i{\sigma _1}/\omega }}{{{\mu _1}}}} (E_{zn}^{(1)}({R_1}) - {E_{szn}}({R_1}))},
				\end{array}} \right.}
				\label{eq:all-1}
			\end{equation}

			\begin{equation}
				\left\{ {\begin{array}{*{20}{c}}
					{{J_n}({k_0}{R_2})\sum\limits_{m =  - \infty }^\infty  {[a_m^3{H_{m - n}^{(2)}}({k_0}{r_{02}}){e^{ - i(n - m){\phi _2}}}]}  + b_n^3{H_n^{(2)}}({k_0}{R_2}) = E_{zn}^{(2)}({R_2})} ,\\
					{{J'_n}({k_0}{R_2})\sum\limits_{m =  - \infty }^\infty  {[a_m^3{H_{m - n}^{(2)}}({k_0}{r_{02}}){e^{ - i(n - m){\phi _2}}}]}  + b_n^3{H_n^{(2)}}'({k_0}{R_2}) = i\omega \sqrt {\frac{{{\varepsilon _2} - i{\sigma _2}/\omega }}{{{\mu _1}}}} E_{zn}^{(2)}({R_2})},
				\end{array}} \right.
				\label{eq:all-2}
			\end{equation}

			\begin{equation}
				\small{
					\left\{ {\begin{array}{*{20}{c}}
						{d_m^2{J_m}({k_0}\eta {r_1}) + g_m^2{H_m^{(2)}}({k_0}\eta {r_1}) = a_m^1{J_m}({k_0}\eta {r_1}) + {H_m^{(2)}}({k_0}\eta {r_1})\sum\limits_{n =  - \infty }^\infty  {[b_n^1{J_{m - n}}({k_0}\eta {r_{01}}){e^{ - i(m - n){\phi _1}}}]} }, \\
						{d_m^2{J'_m}({k_0}\eta {r_1}) + g_m^2{H_m^{(2)}}'({k_0}\eta {r_1}) = a_m^1{J'_m}({k_0}\eta {r_1}) + {H_m^{(2)}}'({k_0}\eta {r_1})\sum\limits_{n =  - \infty }^\infty  {[b_n^1{J_{m - n}}({k_0}\eta {r_{01}}){e^{ - i(m - n){\phi _1}}}]} },
					\end{array}} \right.}
					\label{eq:all-2a}
				\end{equation}
				\begin{equation}
					\small{\left\{ {\begin{array}{*{20}{c}}
						{d_m^2{J_m}({k_0}{r_2}) + g_m^2{H_m^{(2)}}({k_0}{r_2}) = a_m^3{H_m^{(2)}}({k_0}{r_2}) + {J_m}({k_0}{r_2})\sum\limits_{n =  - \infty }^\infty  {[b_n^3{H_{m - n}^{(2)}}({k_0}{r_{02}}){e^{ - i(m - n){\phi _2}}}]} },\\
						{d_m^2{J'_m}({k_0}{r_2}) + g_m^2{H_m^{(2)}}'({k_0}{r_2}) = a_m^3{H_m^{(2)}}'({k_0}{r_2}) + {J'_m}({k_0}{r_2})\sum\limits_{n =  - \infty }^\infty  {[b_n^3{H_{m - n}^{(2)}}({k_0}{r_{02}}){e^{ - i(m - n){\phi _2}}} ]} }.
					\end{array}} \right.}
					\label{eq:all-2b}
				\end{equation}

				\label{eq:all}
			\end{subequations}
		\end{widetext}
Here, we treat the emitter as conducting wire and set its center at points B1 in section 1; thus, we have $d_{01}=||\overrightarrow {{\rho_{OB1}}}| Cos(\phi_{OB1})|$,$r_{01}=|\overrightarrow {{\rho_{OB1}}} |$, $\phi_1=\phi_{OB1}$. Conducting wire has radius $R_1$, conductivity $\sigma_1$ and an n-th order applied voltage of $E_{szn}$(in this article, this parameter appears as the voltage per meter in the 2D model); moreover, we add a variable $E_{zn}^{(1)}$to denote the electric field close to an emitter wire. For the receiver in section 3 (or outside section 3), we have the corresponding constants $d_{02},r_{02},\phi_2,R_2$, and the $\sigma_2$ and the variable $E_{zn}^{(2)}$.
Besides, the time-varying voltage applied to the conducting wire of the emitter is $E_{szn}$; we set a non-zero value for this voltage only at zeroth order, thereby obtaining $E_{szn}=E_{sz}\delta_{0,n}$. Similar as deduction of equation (\ref{eq:simpleModel-all}) we will have
\begin{equation}
	\left\{ {\begin{array}{*{20}{c}}
	{a_m^1 = \sum\limits_{n =  - \infty }^\infty  {[b_n^3{H_{m - n}^{(2)}}({k_0}{r_{02}}){e^{ - i(m - n){\phi _2}}}]} } ,\\
	{a_m^3 = \sum\limits_{n =  - \infty }^\infty  {[b_n^1{J_{m - n}}({k_0}\eta {r_{01}}){e^{ - i(m - n){\phi _1}}}]} },
		\end{array}} \right.
	\label{eq:all-3}
\end{equation}
where $m=0,\pm 1, \pm 2, \dots$. For each order n, we can simplify these eight equations into four equations and thus have only four unknown coefficients.
It should be emphasized the deduction (\ref{eq:all-3}) could be applied for receiver in section 3a and 3b or out of section 3.
\subsection{Materialized model for ss-PCW}
We have previously mentioned that the materialized model for ss-shell covered conducting wire. By applying TO method we could also provide  the materialized model here for ss-PCW.
The schematic of ss-PCW's materialized model is shown in Figure. \ref{fig:sspcwschematic}(b), the electric field in the materialized model consists of two components:
\begin{eqnarray}
				{E_z}({\rho _a},{\phi _a}) = &&\sum\limits_{m =  - \infty }^\infty  {b_m^3{H_m^{(2)}}({k_0}|\overrightarrow {\rho}-\overrightarrow {\rho _{OB2}} |){e^{im{\phi _{B2}}}}  } \nonumber\\
 &&+ \sum\limits_{n =  - \infty }^\infty  {b_n^1{H_n^{(2)}}({k_0}|\overrightarrow {\rho}-\overrightarrow {\rho _{OB1}} |){e^{in{\phi _{B1}}}}  }.	
				\label{eq:wpt-image}
\end{eqnarray}
These two components are expressed in two different coordinate systems with different origin points corresponding to the center points of each wire.

However, it should be emphasized that when the receiver wire is placed in section 3, the TO explanation will lose its utility because the domain (section 3) used for the complementary ss-shell (section 2) is not in vacuum (i.e., the conditions for complementary media are not satisfied). This phenomenon has also been noted in Ref.\cite{luo2009conceal}. Fortunately, under certain specific circumstances, our materialized model will still be applicable even when the complementary media requirements do not hold. We demonstrate this as follows.

When the receiver wire is outside of section 3a ($r_{02}=\rho_{OB_2} >\rho_{OB_1}={r'_{01}}$), we can write the emitter wire's boundary conditions as
\begin{widetext}
	\begin{subequations}
		\begin{equation}
			{{J_n}({k_0}{R'_1}) \sum\limits_{m =  - \infty }^\infty [ {a_m^1{J_{m - n}}({k_0}r'_{01}){e^{ - i(n - m){\phi '_1}}} } ] + b_n^1{H_n^{(2)}}({k_0}{R'_1}) = E_{zn}^1({R'_1})},
		\label{eq:wpt-image-eboundary1}
		\end{equation}
		\begin{equation}
		\small{
			{J'_n({k_0}R'_1) \sum\limits_{m =  - \infty }^\infty [ {a_m^1{J_{m - n}}({k_0}r'_{01}){e^{ - i(n - m){\phi '_1}}}} ] + b_n^1 {H_n^{(2)}}'({k_0}R'_1) = i\omega \sqrt {\frac{{{\varepsilon _1} - i{\sigma '_1}/ \omega }}{{{\mu _1}}}} (E_{zn}^1(R'_1) - {E_{szn}(R'_1)})}
		},
		\label{eq:wpt-image-eboundary2}
		\end{equation}
		\begin{equation}
		{\hbox{where}\;\;}	{a_m^1 = \sum\limits_{k =  - \infty }^\infty [ {b_k^3H_{m - k}^{(2)}({k_0}{r_{02}}){e^{ - i(m - k){\phi _2}}}}   ], \quad\quad ( {r_{02}} > {r'_{01}})},
		\label{eq:wpt-image-eboundary3}
		\end{equation}
		\label{eq:wpt-image-eboundary}
	\end{subequations}
\end{widetext}
where $n=0, \pm 1, \pm 2, \dots$, and the parameters are ${R'_1}=\eta R_1$ (the radius of the emitter wire in the materialized model), $\rho_{OB_1}={r'_{01}}=\eta r_{01}<r_{02}$, and ${\phi '_1}=\phi_1$. Similar to the deductions presented above, we set $\sigma '$ as in equation (\ref{eq:amp-conductivity}), and choose the same $E_{szn}$ as in equations (\ref{eq:all-1}). Finally, we find that equations (\ref{eq:wpt-image-eboundary1}) and (\ref{eq:wpt-image-eboundary2}) are identical to (\ref{eq:all-1}) and (\ref{eq:all-3}). We should note that the relations of equations (\ref{eq:wpt-image-eboundary3}) are satisfied given the condition $r_{02}>{r'_{01}}$. When $r_{02}<{r'_{01}}$, as shown in equations (\ref{eq:waveFunctionTransform-All}), the translation for equations (\ref{eq:wpt-image-eboundary}) will be different. Which also means that the receiver should be outside of section 3b when we consider the equivalent materialized model.

\subsection{Limitation and Extension of TO}
\begin{figure}
	\includegraphics[width=0.5\textwidth]{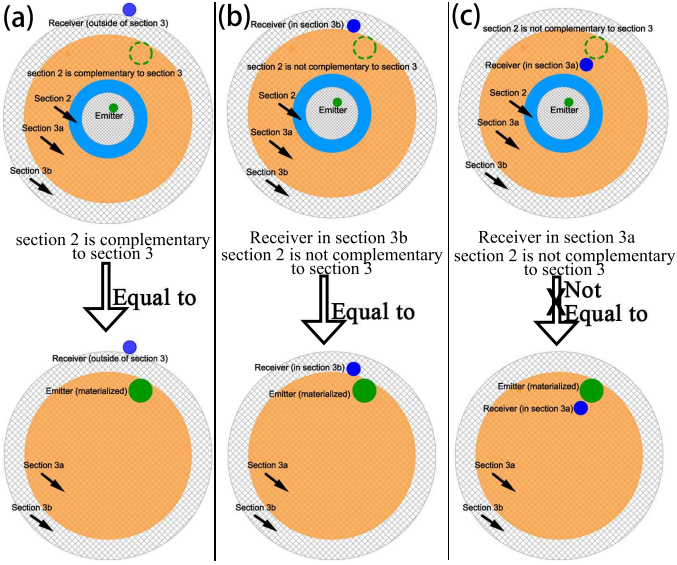}		
	\caption{comparison for ss-PCW and its materialized model: (a), when receiver are out of section 3, complementary media is not destroyed, the TO will give a direct explanation for the equivalence between these two models(the conductivity of materialized emitter in materialized model should be translated as we deduced). (b), receiver situated in section 3b, the complementary media is destroyed, however our demonstration indicate that the equivalence still exist. (c), receiver penetrated in section 3b, the equivalence is invalid.}
	\label{fig:equivalence}
\end{figure}

In conclusion, when the receiver wire penetrates into section 3b (as shown in Figure.\ref{fig:sspcwschematic}(a)), the TO explanation (complementary media) will not function correctly, but the equations for the coefficients in ss-PCW model will remain equivalent to its materialized model (shown in Figure.\ref{fig:sspcwschematic}(b)), which also means that equations (\ref{eq:wpt-image-eboundary}) are still applicable. The relationship between TO, ss-PCW and it's materialized model are presented in Figure.\ref{fig:equivalence}

It is interesting to consider the case in which the receiver wire is moved even closer to the emitter, for example, penetrating into section 3a or the amplified image of the emitter wire's boundary. Where, equation (\ref{eq:wpt-image-eboundary3}) will not be satisfied, and thus, there is no simple geometric explanation as our materialized model based on TO. To calculate the EM fields, we must solve equations (\ref{eq:all}) of emitter and those of receiver numerically.

\subsection{verify the conclusion numerically}
Analytical demonstration is presented above, here we will verify our conclusion numerically through solving equation (\ref{eq:all}) or applying FEM in COMSOL. Once we obtain the EM fields in the model, the total power introduced to the system, transfer power and transfer efficiency are calculated as follows
\begin{subequations}
         \begin{equation}
             {{P_{\hbox{total}}} = \hbox{Re}[\frac{1}{2}E_{sz}^{\rm{*}}(E_{z0}^1 -
{E_{sz}})2\pi {R_1}\sqrt {\frac{{{\varepsilon _c} - i\frac{{{\sigma
_1}}}{\omega }}}{{{\mu _c}}}} \frac{1}{{i\omega {\mu _c}}}]},
         \label{eq:ptotal}
         \end{equation}
         \begin{equation}
             {{P_{\hbox{receiver}}} = \sum\limits_{n =  - \infty }^\infty
{\hbox{Re}[\frac{1}{2}E_{zn}^{{\rm{2*}}}E_{zn}^22\pi {R_2}\sqrt
{\frac{{{\varepsilon _c} - i\frac{{{\sigma _2}}}{\omega }}}{{{\mu _c}}}}
\frac{1}{{i\omega {\mu _c}}}} ]},
         \label{eq:preceiver}
         \end{equation}
         \begin{equation}
             {\zeta  = \frac{{{P_{\hbox{receiver}}}}}{{{P_{\hbox{total}}}}}}.
         \label{eq:ptpr}
         \end{equation}
         \label{eq:efficiency}
     \end{subequations}

As presented in the above demonstration, we have already demonstrate the TO based materialized could be equivalent to its ss-PCW even when section 2 is not complementary to section 3. As the equivalence also lead to the equivalent transfer power and efficiency, we could verify our demonstration numerically by using the FEM (COMSOL) and our series expansion solution. We present the numerical results in two different scenarios, in which receiver is situated in section 3b and section 3a respectively.

\begin{figure}
	\includegraphics[width=0.5\textwidth]{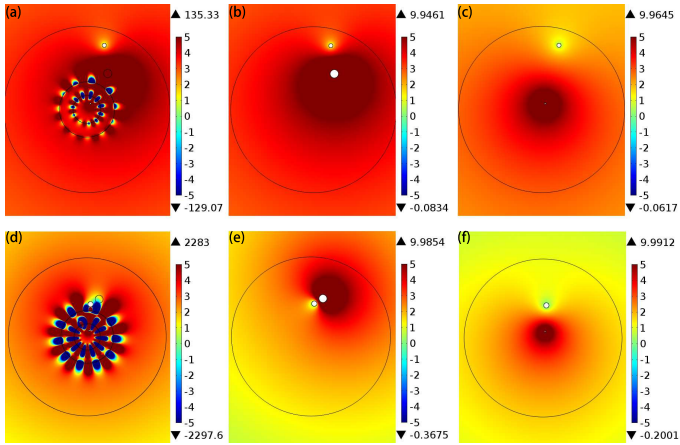}		
	\caption{Electric fields (z component) for ss-PCW 1 \& 2: (a),electric fields (z component) in ss-PCW 1. (b),electric fields (z component) in materialized model of ss-PCW 1. (c),ordinary PCW. Compare (a) with (b), we will find the equivalence between ss-PCW 1 and its materialized model. (d),electric fields (z component) in ss-PCW 2. (e),electric fields (z component) in materialized model of ss-PCW 2. (f),electric fields (z component) in ordinary PCW. Compare (d) with (e), we will find there isn't equivalence between ss-PCW 2 and its materialized model.}
	\label{fig:sspcw}
\end{figure}

First, we consider the case in which the receiver wire is located in section 3b, as shown in Figure.\ref{fig:equivalence}(b), which means that $\eta (r_{01}+R_1)< r_{02}$. Here we calculate the model for $r_{01}=0.1$ m, $d_{01}=0.1$ m, $R_{1}=0.01$ m, $R_{2}=0.03$ m, $d_{02}=0.5$ m, and $r_{02}=0.96$ m, and set the amplification factor to $\eta = 6$ (we call this scenario ss-PCW 1). Because  $\eta (r_{01}+R_1)< r_{02}$, we can consider a equivalent materialized model with only receiver and emitter wire with parameters of  $r'_{01}=0.6$ m, $d'_{01}=0.6$ m, $R'_{1}=0.06$ m, $R_{2}=0.03$ m, $d_{02}=0.5$ m and $r_{02}=0.96$ m. The voltage applied to the emitter wire is $10$ V/m (and the wavelength is $100$ m). The material of the emitter wire is conducting (with $\sigma=5\times 10^5 $ S/m), and the receiver wire is connected to an external load ($0.6+0.6i$$\Omega$/m). The results are listed in Table \ref{tab:result1}. As we expected, although the TO approach loses its validity when the receiver wire enter section 3b, the materialized model remains useful.
Figure.\ref{fig:sspcw}(a),(b),(c) shows the electric fields (z component) in the three simulated models. By analyzing the distribution of $E_z$, we will find the equivalence between ss-PCW and its materialized model easily.
\begin{table*}
			\caption{\label{tab:result1}Results from COMSOL and the series expansion solution for ss-PCW 1}
			\begin{ruledtabular}
				\begin{tabular}{|c|c|c|c||c|c|c|}
					&\multicolumn{3}{c||}{Series expansion solution}&\multicolumn{3}{c|}{COMSOL results}\\ \hline
					 &\multicolumn{2}{c|}{Power($W/m$)}&Ratio(\%)&\multicolumn{2}{c|}{Power($W/m$)}&Ratio(\%)\\
					\hline
					&Receiver&Emitter&Efficiency&Receiver&Emitter&Efficiency\\ \hline
					ss-PCW&$0.227$&$0.559$&$40.6\%$&$0.223$&$0.538$&$41.4\%$\\
					materialized model&$0.227$&$0.559$&$40.6\%$&$0.225$&$0.543$&$41.4\%$\\
					ordinary PCW&$0.057$&$0.229$&$24.7\%$&$0.056$&$0.221$&$25.3\%$\\
				\end{tabular}
			\end{ruledtabular}
\end{table*}

\begin{table*}
			\caption{\label{tab:result2}Results from COMSOL and the series expansion solution for ss-PCW 2}
			\begin{ruledtabular}
				\begin{tabular}{|c|c|c|c||c|c|c|}
					&\multicolumn{3}{c||}{Series expansion solution}&\multicolumn{3}{c|}{COMSOL results}\\ \hline
					 &\multicolumn{2}{c|}{Power($W/m$)}&Ratio(\%)&\multicolumn{2}{c|}{Power($W/m$)}&Ratio(\%)\\
					\hline
					&Receiver&Emitter&Efficiency&Receiver&Emitter&Efficiency\\ \hline
					ss-PCW&$3.82$&$4.39$&$87.0\%$&$3.67$&$4.27$&$86.0\%$\\
					materialized model&$0.368$&$0.853$&$43.1\%$&$0.368$&$0.851$&$43.2\%$\\
					ordinary PCW&$0.021$&$0.235$&$9.1\%$&$0.021$&$0.235$&$9.1\%$\\

				\end{tabular}
			\end{ruledtabular}
		\end{table*}

Similar as the above presented comparison, when receiver wire penetrated into section 3a (we call it ss-PCW 2, in fact ss-PCW 2 can not be represented by its related materialized model as we demonstrated analytically above). Based on our series expansion solution and COMSOL results we could also give the comparison as Table \ref{tab:result2}. In the ss-PCW 2 scenario, we set the parameters as $r_{01}=0.5$ m, $d_{01}=0.3$ m, $R_{1}=0.05,R_{2}=0.2$ m, $d_{02}=0.5$ m and $r_{02}=2.49$ m, with amplification factor as $\eta=6$, which satisfies the condition $\eta(r_{01}+R1)>r_{02}$. The wire's material and external load are taken to be the same as in ss-PCW 1, so as the applied voltage.Figure.\ref{fig:sspcw}(d),(e),(f) shows the electric fields (z component) for ss-PCW 2 scenario.

In conclusion, it is obvious that the scenario ss-PCW 2 cannot be explained in terms of the TO approach and also cannot be correctly represented by a corresponding materialized model, while ss-PCW 1 shows the TO based materialized model still be applicable even the condition of complementary media is destroyed.

Moreover, it should be noted, because of non-monotonic transforms, the computational error in simulation for superscatterers from FEM is hard to control, this phenomenon is also discussed in Ref.\cite{aznavourian2014morphing}. To give the above mentioned meaningful results the computational time for FEM is much longer than  that cost by our methodology. Besides, FEM requires an increasing number of mesh elements as the scale of the model grows, the time required for compuation and the iterative error will increase even if there is sufficient memory. Our results obtained from series expansion solution are more reliable for simulation of super scatterer, and need less computational time, especially for large-scale model.

\section{Conclusion and Discussion}
In this article we extend the TO explanation in metamaterial modified WPT system, and from rigorous scattering analysis, we establish a equivalent model for the system, named materialized model. From our analytical analysis, we demonstrate that the TO based materialized model could be applied even when TO approach could not give a direct explanation, where the requirements for complementary are not satisfied. Besides, we present the numerical results from FEM and our series expansion solution to verify our findings. The conclusion is firmly verified by these numerical results.

The wireless power transfer system here is simplified as two parallel conductor wires, however, the results and methodology should also be useful for a more complicated model, such as more pairs of wires or coils, and even a 3D helix coil.

Moreover, our methodology which is based on multiple scattering theory could deal with the calculation of EM fields in superscatterer modified WPT more efficiently than FEM, especially in large-scale model.

Furthermore, our analytical demonstration is not under some certain frequency, which means the results could be applied in researches under a broad range of wavelength, e.g. the research in active cloak and antenna. The demonstration was under impedance boundary condition, which could be applied in many researches. We expect the analysing for a more specific boundary condition like perfect electric conductor boundary could reveal more special properties by applying our methodology. Besides, the TO explanation for WPT is very important, which might significantly improve the WPT research.


\section*{Acknowledgments}
\indent This work was sponsored in part by the National Natural Science Foundation of China under Grants No. 51277120, and by the International S\&T Cooperation Program of China (2012DFG01790).

\bibliography{extendTOcitation}

\providecommand{\noopsort}[1]{}\providecommand{\singleletter}[1]{#1}%
\begin{thebibliography}{22}%
\makeatletter
\providecommand \@ifxundefined [1]{%
 \@ifx{#1\undefined}
}%
\providecommand \@ifnum [1]{%
 \ifnum #1\expandafter \@firstoftwo
 \else \expandafter \@secondoftwo
 \fi
}%
\providecommand \@ifx [1]{%
 \ifx #1\expandafter \@firstoftwo
 \else \expandafter \@secondoftwo
 \fi
}%
\providecommand \natexlab [1]{#1}%
\providecommand \enquote  [1]{``#1''}%
\providecommand \bibnamefont  [1]{#1}%
\providecommand \bibfnamefont [1]{#1}%
\providecommand \citenamefont [1]{#1}%
\providecommand \href@noop [0]{\@secondoftwo}%
\providecommand \href [0]{\begingroup \@sanitize@url \@href}%
\providecommand \@href[1]{\@@startlink{#1}\@@href}%
\providecommand \@@href[1]{\endgroup#1\@@endlink}%
\providecommand \@sanitize@url [0]{\catcode `\\12\catcode `\$12\catcode
  `\&12\catcode `\#12\catcode `\^12\catcode `\_12\catcode `\%12\relax}%
\providecommand \@@startlink[1]{}%
\providecommand \@@endlink[0]{}%
\providecommand \url  [0]{\begingroup\@sanitize@url \@url }%
\providecommand \@url [1]{\endgroup\@href {#1}{\urlprefix }}%
\providecommand \urlprefix  [0]{URL }%
\providecommand \Eprint [0]{\href }%
\providecommand \doibase [0]{http://dx.doi.org/}%
\providecommand \selectlanguage [0]{\@gobble}%
\providecommand \bibinfo  [0]{\@secondoftwo}%
\providecommand \bibfield  [0]{\@secondoftwo}%
\providecommand \translation [1]{[#1]}%
\providecommand \BibitemOpen [0]{}%
\providecommand \bibitemStop [0]{}%
\providecommand \bibitemNoStop [0]{.\EOS\space}%
\providecommand \EOS [0]{\spacefactor3000\relax}%
\providecommand \BibitemShut  [1]{\csname bibitem#1\endcsname}%
\let\auto@bib@innerbib\@empty
\bibitem [{\citenamefont {Sihvola}(2007)}]{sihvola2007metamaterials}%
  \BibitemOpen
  \bibfield  {author} {\bibinfo {author} {\bibfnamefont {A.}~\bibnamefont
  {Sihvola}},\ }\href@noop {} {\bibfield  {journal} {\bibinfo  {journal}
  {Metamaterials}\ }\textbf {\bibinfo {volume} {1}},\ \bibinfo {pages} {2}
  (\bibinfo {year} {2007})}\BibitemShut {NoStop}%
\bibitem [{\citenamefont {Shamonina}\ and\ \citenamefont
  {Solymar}(2007)}]{shamonina2007metamaterials}%
  \BibitemOpen
  \bibfield  {author} {\bibinfo {author} {\bibfnamefont {E.}~\bibnamefont
  {Shamonina}}\ and\ \bibinfo {author} {\bibfnamefont {L.}~\bibnamefont
  {Solymar}},\ }\href@noop {} {\bibfield  {journal} {\bibinfo  {journal}
  {Metamaterials}\ }\textbf {\bibinfo {volume} {1}},\ \bibinfo {pages} {12}
  (\bibinfo {year} {2007})}\BibitemShut {NoStop}%
\bibitem [{\citenamefont {Pendry}(2000)}]{pendry2000perfectlens}%
  \BibitemOpen
  \bibfield  {author} {\bibinfo {author} {\bibfnamefont {J.~B.}\ \bibnamefont
  {Pendry}},\ }\href@noop {} {\bibfield  {journal} {\bibinfo  {journal} {Phys.
  Rev. Lett.}\ }\textbf {\bibinfo {volume} {85}},\ \bibinfo {pages} {3966}
  (\bibinfo {year} {2000})}\BibitemShut {NoStop}%
\bibitem [{\citenamefont {Urzhumov}\ and\ \citenamefont
  {Smith}(2011)}]{metaWPT}%
  \BibitemOpen
  \bibfield  {author} {\bibinfo {author} {\bibfnamefont {Y.}~\bibnamefont
  {Urzhumov}}\ and\ \bibinfo {author} {\bibfnamefont {D.~R.}\ \bibnamefont
  {Smith}},\ }\href@noop {} {\bibfield  {journal} {\bibinfo  {journal}
  {Physical Review B}\ }\textbf {\bibinfo {volume} {83}},\ \bibinfo {pages}
  {205114} (\bibinfo {year} {2011})}\BibitemShut {NoStop}%
\bibitem [{\citenamefont {Kong}\ \emph {et~al.}(2007)\citenamefont {Kong},
  \citenamefont {Wu}, \citenamefont {Kong}, \citenamefont {Huangfu},
  \citenamefont {Xi},\ and\ \citenamefont {Chen}}]{antenna}%
  \BibitemOpen
  \bibfield  {author} {\bibinfo {author} {\bibfnamefont {F.}~\bibnamefont
  {Kong}}, \bibinfo {author} {\bibfnamefont {B.-I.}\ \bibnamefont {Wu}},
  \bibinfo {author} {\bibfnamefont {J.~A.}\ \bibnamefont {Kong}}, \bibinfo
  {author} {\bibfnamefont {J.}~\bibnamefont {Huangfu}}, \bibinfo {author}
  {\bibfnamefont {S.}~\bibnamefont {Xi}}, \ and\ \bibinfo {author}
  {\bibfnamefont {H.}~\bibnamefont {Chen}},\ }\href@noop {} {\bibfield
  {journal} {\bibinfo  {journal} {Applied Physics Letters}\ }\textbf {\bibinfo
  {volume} {91}},\  (\bibinfo {year} {2007})}\BibitemShut {NoStop}%
\bibitem [{\citenamefont {Wang}\ \emph {et~al.}(2011)\citenamefont {Wang},
  \citenamefont {Teo}, \citenamefont {Nishino}, \citenamefont {Yerazunis},
  \citenamefont {Barnwell},\ and\ \citenamefont {Zhang}}]{experimentWPTmeta}%
  \BibitemOpen
  \bibfield  {author} {\bibinfo {author} {\bibfnamefont {B.}~\bibnamefont
  {Wang}}, \bibinfo {author} {\bibfnamefont {K.~H.}\ \bibnamefont {Teo}},
  \bibinfo {author} {\bibfnamefont {T.}~\bibnamefont {Nishino}}, \bibinfo
  {author} {\bibfnamefont {W.}~\bibnamefont {Yerazunis}}, \bibinfo {author}
  {\bibfnamefont {J.}~\bibnamefont {Barnwell}}, \ and\ \bibinfo {author}
  {\bibfnamefont {J.}~\bibnamefont {Zhang}},\ }\href@noop {} {\bibfield
  {journal} {\bibinfo  {journal} {Applied Physics Letters}\ }\textbf {\bibinfo
  {volume} {98}},\  (\bibinfo {year} {2011})}\BibitemShut {NoStop}%
\bibitem [{\citenamefont {Leonhardt}(2006)}]{tranU}%
  \BibitemOpen
  \bibfield  {author} {\bibinfo {author} {\bibfnamefont {U.}~\bibnamefont
  {Leonhardt}},\ }\href@noop {} {\bibfield  {journal} {\bibinfo  {journal}
  {Science}\ }\textbf {\bibinfo {volume} {312}},\ \bibinfo {pages} {1777}
  (\bibinfo {year} {2006})}\BibitemShut {NoStop}%
\bibitem [{\citenamefont {Pendry}\ \emph {et~al.}(2006)\citenamefont {Pendry},
  \citenamefont {Schurig},\ and\ \citenamefont
  {Smith}}]{pendry2006controlling}%
  \BibitemOpen
  \bibfield  {author} {\bibinfo {author} {\bibfnamefont {J.~B.}\ \bibnamefont
  {Pendry}}, \bibinfo {author} {\bibfnamefont {D.}~\bibnamefont {Schurig}}, \
  and\ \bibinfo {author} {\bibfnamefont {D.~R.}\ \bibnamefont {Smith}},\
  }\href@noop {} {\bibfield  {journal} {\bibinfo  {journal} {science}\ }\textbf
  {\bibinfo {volume} {312}},\ \bibinfo {pages} {1780} (\bibinfo {year}
  {2006})}\BibitemShut {NoStop}%
\bibitem [{\citenamefont {Chen}\ \emph {et~al.}(2010)\citenamefont {Chen},
  \citenamefont {Chan},\ and\ \citenamefont {Sheng}}]{chen2010transformation}%
  \BibitemOpen
  \bibfield  {author} {\bibinfo {author} {\bibfnamefont {H.}~\bibnamefont
  {Chen}}, \bibinfo {author} {\bibfnamefont {C.}~\bibnamefont {Chan}}, \ and\
  \bibinfo {author} {\bibfnamefont {P.}~\bibnamefont {Sheng}},\ }\href@noop {}
  {\bibfield  {journal} {\bibinfo  {journal} {Nature materials}\ }\textbf
  {\bibinfo {volume} {9}},\ \bibinfo {pages} {387} (\bibinfo {year}
  {2010})}\BibitemShut {NoStop}%
\bibitem [{\citenamefont {Huang}\ \emph {et~al.}(2012)\citenamefont {Huang},
  \citenamefont {Urzhumov}, \citenamefont {Smith}, \citenamefont {Hoo~Teo},\
  and\ \citenamefont {Zhang}}]{superlensWPT}%
  \BibitemOpen
  \bibfield  {author} {\bibinfo {author} {\bibfnamefont {D.}~\bibnamefont
  {Huang}}, \bibinfo {author} {\bibfnamefont {Y.}~\bibnamefont {Urzhumov}},
  \bibinfo {author} {\bibfnamefont {D.~R.}\ \bibnamefont {Smith}}, \bibinfo
  {author} {\bibfnamefont {K.}~\bibnamefont {Hoo~Teo}}, \ and\ \bibinfo
  {author} {\bibfnamefont {J.}~\bibnamefont {Zhang}},\ }\href@noop {}
  {\bibfield  {journal} {\bibinfo  {journal} {Journal of Applied Physics}\
  }\textbf {\bibinfo {volume} {111}},\  (\bibinfo {year} {2012})}\BibitemShut
  {NoStop}%
\bibitem [{\citenamefont {Zhang}\ \emph {et~al.}(2008)\citenamefont {Zhang},
  \citenamefont {Chen}, \citenamefont {Wu},\ and\ \citenamefont
  {Kong}}]{PhysRevLett.100.063904active}%
  \BibitemOpen
  \bibfield  {author} {\bibinfo {author} {\bibfnamefont {B.}~\bibnamefont
  {Zhang}}, \bibinfo {author} {\bibfnamefont {H.}~\bibnamefont {Chen}},
  \bibinfo {author} {\bibfnamefont {B.-I.}\ \bibnamefont {Wu}}, \ and\ \bibinfo
  {author} {\bibfnamefont {J.~A.}\ \bibnamefont {Kong}},\ }\href {\doibase
  10.1103/PhysRevLett.100.063904} {\bibfield  {journal} {\bibinfo  {journal}
  {Phys. Rev. Lett.}\ }\textbf {\bibinfo {volume} {100}},\ \bibinfo {pages}
  {063904} (\bibinfo {year} {2008})}\BibitemShut {NoStop}%
\bibitem [{\citenamefont {Selvanayagam}\ and\ \citenamefont
  {Eleftheriades}(2012)}]{ieee6330979active}%
  \BibitemOpen
  \bibfield  {author} {\bibinfo {author} {\bibfnamefont {M.}~\bibnamefont
  {Selvanayagam}}\ and\ \bibinfo {author} {\bibfnamefont {G.}~\bibnamefont
  {Eleftheriades}},\ }\href@noop {} {\bibfield  {journal} {\bibinfo  {journal}
  {Antennas and Wireless Propagation Letters, IEEE}\ }\textbf {\bibinfo
  {volume} {11}},\ \bibinfo {pages} {1226} (\bibinfo {year}
  {2012})}\BibitemShut {NoStop}%
\bibitem [{\citenamefont {Selvanayagam}\ and\ \citenamefont
  {Eleftheriades}(2013)}]{PhysRevX.3.041011active}%
  \BibitemOpen
  \bibfield  {author} {\bibinfo {author} {\bibfnamefont {M.}~\bibnamefont
  {Selvanayagam}}\ and\ \bibinfo {author} {\bibfnamefont {G.~V.}\ \bibnamefont
  {Eleftheriades}},\ }\href {\doibase 10.1103/PhysRevX.3.041011} {\bibfield
  {journal} {\bibinfo  {journal} {Phys. Rev. X}\ }\textbf {\bibinfo {volume}
  {3}},\ \bibinfo {pages} {041011} (\bibinfo {year} {2013})}\BibitemShut
  {NoStop}%
\bibitem [{\citenamefont {Yang}\ \emph {et~al.}(2008)\citenamefont {Yang},
  \citenamefont {Chen}, \citenamefont {Luo},\ and\ \citenamefont
  {Ma}}]{yang2008superscatterer}%
  \BibitemOpen
  \bibfield  {author} {\bibinfo {author} {\bibfnamefont {T.}~\bibnamefont
  {Yang}}, \bibinfo {author} {\bibfnamefont {H.}~\bibnamefont {Chen}}, \bibinfo
  {author} {\bibfnamefont {X.}~\bibnamefont {Luo}}, \ and\ \bibinfo {author}
  {\bibfnamefont {H.}~\bibnamefont {Ma}},\ }\href@noop {} {\bibfield  {journal}
  {\bibinfo  {journal} {Optics express}\ }\textbf {\bibinfo {volume} {16}},\
  \bibinfo {pages} {18545} (\bibinfo {year} {2008})}\BibitemShut {NoStop}%
\bibitem [{\citenamefont {Luo}\ \emph {et~al.}(2009)\citenamefont {Luo},
  \citenamefont {Yang}, \citenamefont {Gu}, \citenamefont {Chen},\ and\
  \citenamefont {Ma}}]{luo2009conceal}%
  \BibitemOpen
  \bibfield  {author} {\bibinfo {author} {\bibfnamefont {X.}~\bibnamefont
  {Luo}}, \bibinfo {author} {\bibfnamefont {T.}~\bibnamefont {Yang}}, \bibinfo
  {author} {\bibfnamefont {Y.}~\bibnamefont {Gu}}, \bibinfo {author}
  {\bibfnamefont {H.}~\bibnamefont {Chen}}, \ and\ \bibinfo {author}
  {\bibfnamefont {H.}~\bibnamefont {Ma}},\ }\href@noop {} {\bibfield  {journal}
  {\bibinfo  {journal} {Applied physics letters}\ }\textbf {\bibinfo {volume}
  {94}},\ \bibinfo {pages} {223513} (\bibinfo {year} {2009})}\BibitemShut
  {NoStop}%
\bibitem [{\citenamefont {Ramahi}\ \emph {et~al.}(2012)\citenamefont {Ramahi},
  \citenamefont {Almoneef}, \citenamefont {AlShareef},\ and\ \citenamefont
  {Boybay}}]{metaharvesting}%
  \BibitemOpen
  \bibfield  {author} {\bibinfo {author} {\bibfnamefont {O.~M.}\ \bibnamefont
  {Ramahi}}, \bibinfo {author} {\bibfnamefont {T.~S.}\ \bibnamefont
  {Almoneef}}, \bibinfo {author} {\bibfnamefont {M.}~\bibnamefont {AlShareef}},
  \ and\ \bibinfo {author} {\bibfnamefont {M.~S.}\ \bibnamefont {Boybay}},\
  }\href@noop {} {\bibfield  {journal} {\bibinfo  {journal} {Applied Physics
  Letters}\ }\textbf {\bibinfo {volume} {101}},\  (\bibinfo {year}
  {2012})}\BibitemShut {NoStop}%
\bibitem [{\citenamefont {Rajagopalan}\ \emph {et~al.}(2014)\citenamefont
  {Rajagopalan}, \citenamefont {RamRakhyani}, \citenamefont {Schurig},\ and\
  \citenamefont {Lazzi}}]{compactShort}%
  \BibitemOpen
  \bibfield  {author} {\bibinfo {author} {\bibfnamefont {A.}~\bibnamefont
  {Rajagopalan}}, \bibinfo {author} {\bibfnamefont {A.~K.}\ \bibnamefont
  {RamRakhyani}}, \bibinfo {author} {\bibfnamefont {D.}~\bibnamefont
  {Schurig}}, \ and\ \bibinfo {author} {\bibfnamefont {G.}~\bibnamefont
  {Lazzi}},\ }\href@noop {} {\bibfield  {journal} {\bibinfo  {journal} {IEEE
  TRANSACTIONS ON MICROWAVE THEORY AND TECHNIQUES}\ }\textbf {\bibinfo {volume}
  {62}},\ \bibinfo {pages} {947} (\bibinfo {year} {2014})}\BibitemShut
  {NoStop}%
\bibitem [{\citenamefont {Pendry}\ and\ \citenamefont
  {Ramakrishna}(2003)}]{pendry2003complementary}%
  \BibitemOpen
  \bibfield  {author} {\bibinfo {author} {\bibfnamefont {J.~B.}\ \bibnamefont
  {Pendry}}\ and\ \bibinfo {author} {\bibfnamefont {S.~A.}\ \bibnamefont
  {Ramakrishna}},\ }\href@noop {} {\bibfield  {journal} {\bibinfo  {journal}
  {Journal of Physics: Condensed Matter}\ }\textbf {\bibinfo {volume} {15}},\
  \bibinfo {pages} {6345} (\bibinfo {year} {2003})}\BibitemShut {NoStop}%
\bibitem [{\citenamefont {Yuferev}\ and\ \citenamefont
  {Ida}(1999)}]{yuferev1999selection}%
  \BibitemOpen
  \bibfield  {author} {\bibinfo {author} {\bibfnamefont {S.}~\bibnamefont
  {Yuferev}}\ and\ \bibinfo {author} {\bibfnamefont {N.}~\bibnamefont {Ida}},\
  }\href@noop {} {\bibfield  {journal} {\bibinfo  {journal} {Magnetics, IEEE
  Transactions on}\ }\textbf {\bibinfo {volume} {35}},\ \bibinfo {pages} {1486}
  (\bibinfo {year} {1999})}\BibitemShut {NoStop}%
\bibitem [{\citenamefont {Kurs}\ \emph {et~al.}(2007)\citenamefont {Kurs},
  \citenamefont {Karalis}, \citenamefont {Moffatt}, \citenamefont
  {Joannopoulos}, \citenamefont {Fisher},\ and\ \citenamefont
  {Solja{\v{c}}i{\'c}}}]{kurs2007science}%
  \BibitemOpen
  \bibfield  {author} {\bibinfo {author} {\bibfnamefont {A.}~\bibnamefont
  {Kurs}}, \bibinfo {author} {\bibfnamefont {A.}~\bibnamefont {Karalis}},
  \bibinfo {author} {\bibfnamefont {R.}~\bibnamefont {Moffatt}}, \bibinfo
  {author} {\bibfnamefont {J.~D.}\ \bibnamefont {Joannopoulos}}, \bibinfo
  {author} {\bibfnamefont {P.}~\bibnamefont {Fisher}}, \ and\ \bibinfo {author}
  {\bibfnamefont {M.}~\bibnamefont {Solja{\v{c}}i{\'c}}},\ }\href@noop {}
  {\bibfield  {journal} {\bibinfo  {journal} {science}\ }\textbf {\bibinfo
  {volume} {317}},\ \bibinfo {pages} {83} (\bibinfo {year} {2007})}\BibitemShut
  {NoStop}%
\bibitem [{\citenamefont {Chew}(1995)}]{chew1995waves}%
  \BibitemOpen
  \bibfield  {author} {\bibinfo {author} {\bibfnamefont {W.~C.}\ \bibnamefont
  {Chew}},\ }\href@noop {} {\emph {\bibinfo {title} {Waves and fields in
  inhomogeneous media}}},\ Vol.\ \bibinfo {volume} {522}\ (\bibinfo
  {publisher} {IEEE press New York},\ \bibinfo {year} {1995})\BibitemShut
  {NoStop}%
\bibitem [{\citenamefont {Aznavourian}\ and\ \citenamefont
  {Guenneau}(2014)}]{aznavourian2014morphing}%
  \BibitemOpen
  \bibfield  {author} {\bibinfo {author} {\bibfnamefont {R.}~\bibnamefont
  {Aznavourian}}\ and\ \bibinfo {author} {\bibfnamefont {S.}~\bibnamefont
  {Guenneau}},\ }\href@noop {} {\bibfield  {journal} {\bibinfo  {journal}
  {Optics express}\ }\textbf {\bibinfo {volume} {22}},\ \bibinfo {pages}
  {28301} (\bibinfo {year} {2014})}\BibitemShut {NoStop}%
\end{thebibliography}%
\end{document}